\documentstyle[preprint,aps,prb,epsf]{revtex}
\begin{document}
\preprint
\widetext
\title{Proof of phase separation in the binary-alloy problem:
       the one-dimensional spinless Falicov-Kimball model}
\author{J.\ K.\ Freericks$^1$, Ch. Gruber$^2$, and N. Macris$^2$}
\address{$^1$Department of Physics, Georgetown University, Washington, DC 20057,
U.S.A.\\
$^2$Institut de Physique Th\'eorique, Ecole Polytechnique F\'ed\'erale de 
Lausanne,\\
PHB-Ecublens, CH-1015, Lausanne, Switzerland
\\
}
\date{\today}
\maketitle
\widetext
\begin{abstract}
The ground states of the one-dimensional Falicov-Kimball model are investigated
in the small-coupling limit, using nearly degenerate perturbation theory.
For rational electron and ion densities, respectively equal to $\frac{p}{q}$, 
$\frac{p_i}{q}$, with $p$ relatively prime to $q$ and $\frac{p_i}{q}$ close 
enough to $\frac{1}{2}$, we find that in the ground state the ion configuration 
has period $q$.  The situation is analogous to the Peierls instability 
where the usual arguments predict a period-$q$ state that produces a gap at 
the Fermi level and is insulating.  However for $\frac{p_i}{q}$ far enough 
from $\frac{1}{2}$, this phase becomes unstable against phase separation.  The 
ground state is a mixture of a period-$q$ ionic configuration and an empty (or 
full) configuration, where both configurations have the same electron density 
to leading order.  Combining these new results with those previously obtained 
for strong coupling, it follows that a phase transition occurs in the ground 
state, as a function of the coupling, for ion densities far enough from 
$\frac{1}{2}$.
\end{abstract}
%\renewcommand{\thefootnote}{\copyright}
%\footnotetext{ 1996 by the authors.  Reproduction of this article by any means
%is permitted for non-commercial purposes.}
%\renewcommand{\thefootnote}{\alpha{footnote}}


\section{Introduction}

The theory of the electronic band-structure of solids is one of the oldest
theories in condensed-matter physics, dating back to 1928 when Bloch proved the 
existence of electronic bands in solids\cite{bloch}.  It was soon discovered
that most metals could be described with the {\it nearly free electron model} 
where the
periodic ion potential felt by the electrons is weak.  Initial emphasis
in the development of band theory focused on one-dimensional models where
both transfer-matrix results\cite{kramers} and exact 
solutions\cite{morse,kronig_penney} could be found.  It was only after the
development of the pseudopotential method\cite{kleinman_phillips} in the
1960's that the success of the nearly free electron model was understood.

In the 1950's, Peierls\cite{peierls} reexamined the perturbation theory for the 
nearly
free electron model and found that, in one-dimension, a {\it static} distortion
would always reduce the energy of a solid, because the opening of a gap (at the
Fermi level) in the electronic band-structure would lower the energy of 
the occupied electronic states and raise the energy of the unoccupied states.
Such a distortion would produce an {\it insulator} from the parent {\it metal}
and Peierls's work led to the conclusion that there can never be a
one-dimensional metal.
Fr\"ohlich\cite{frohlich} used nearly degenerate perturbation theory to show
that the decrease in the electronic energy was on the order of 
$\delta^2\ln\delta$ for a periodic distortion of amplitude $\delta$, while the
elastic energy was on the order of $\delta^2$, so the net effect of the 
distortion was to reduce the ground-state energy (if $\delta$ was small enough).
Two chemists, Longuet-Higgins and Salem\cite{longuet}, independently arrived
at the same conclusions by examining a general class of Hamiltonians for
ring-shaped molecules.  Note that these conclusions explicitly neglect the 
quantum fluctuations of the phonons about the distorted state, a feature which
can lead to a stabilization of the undistorted phase, as illustrated in 
recent rigorous work on the one-dimensional Holstein model.\cite{benfatto}

The Peierls distortion is generally studied at half-filling for the electrons,
in which case the distortion leads to a doubling of the unit cell.  The
conventional wisdom is that the {\it lowest} periodic structure that produces
a gap at the Fermi level will be the true ground state, or, in other words,
the Peierls distortion is stable against any higher-order distortions.  This
result has been recently proven to be true at half 
filling\cite{lieb_peierls,lebo_macris}.

In one dimension, the restriction to half-filling is not necessary, since the
same arguments used by Peierls in his original discussion, could be extended
to arbitrary electron filling, with the result that a periodic distortion
that produces a gap at the Fermi level will always lower the total energy
of the system.  Furthermore, one can generalize Peierls' arguments to include
other models, such as models for alloy formation.  Historically, these alloy
models were studied before the models of lattice distortions, where 
Hume-Rothery\cite{hume_rothery} pointed out that certain alloys form only
when the electron-atom ratio fell within very narrow ranges.  In one-dimension,
Peierls-like arguments establish the Hume-Rothery rule, since the periodic
arrangement of the ions will always produce the largest gap at the Fermi level.
In the alloy picture, it is the periodic ordering of the ions that creates
the band structure with a gap at the Fermi level, and it is a reasonable 
assumption
to neglect the quantum fluctuations that would be induced if the ions
became itinerant since the energetics of interchanging two ionic species
should be large in a one-dimensional system.

The simplest model of a binary alloy is the spinless Falicov-Kimball 
model\cite{falicov_kimball} in which itinerant spinless electrons interact
with static ions:
\begin{equation}
H=-t\sum_{i=1}^N (c_i^{\dagger}c_{i+1}+c_{i+1}^{\dagger}c_i) - U 
\sum_{i=1}^N c_i^{\dagger} c_i w_i\quad ,
\label{eq: H_fk}
\end{equation}
where $c_i^{\dagger}$ is the creation operator for an electron at site $i$,
$w_i$ is a classical variable that is 1 if the site is occupied by an A ion
and 0 if the site is occupied by a B ion.  The hopping integral is $t$ and 
$U$ denotes the difference in on-site energies for an electron on a B site
minus that of an electron on an A site.  The electron density, $\rho_e$,
is the number of electrons per site $(\rho_e=\frac{1}{N}
\sum_i<c_i^{\dagger}c_i>)$,
and similarly the ion density $\rho_i$ is the number of A ions per
site ($\rho_i=\frac{1}{N}\sum_iw_i)$ with $N$ the number of lattice sites.  We 
are interested in the thermodynamic limit, and so we take $N\rightarrow\infty$,
but maintain finite values for both $\rho_e$ and $\rho_i$.
In the alloy picture $(U>0)$, 
one can envision that the electrons are donated by one
of the ionic species (say the A ion), in which case a study of the neutral
case (where $\rho_i=\rho_e$) becomes most relevant. 

This model was first 
proposed by Falicov and Kimball\cite{falicov_kimball}  to discuss 
semiconductor-insulator transitions in certain rare-earth oxides and borides.
At zero temperature, the electrons are localized on the sites of the metallic
ions.  As the temperature is increased, itinerant electrons and localized
holes are simultaneously produced by excitation.  The Coulomb interaction
is screened out, except for the on-site hole-hole repulsion, which
is very strong (hard-core repulsion) and the on-site electron-hole
attraction, which was assumed to be responsible for the transition.  In this 
case, $U$ is positive, and the system is neutral $\rho_e=\rho_i$.  Later,
the same model was employed to study ordering of rare-earth ions in
mixed-valence systems\cite{sakurai_schlottmann}.  In this application,
the itinerant electrons are $d$-electrons, and the static particles are
localized $f$-electrons.  The interaction is then repulsive $(U<0)$, and
conservation of the total number of electrons leads to $\rho_e+\rho_i=~
{\rm constant}$, with the constant usually taken to be one.

More recently\cite{lieb_fk}, work
on the Falicov-Kimball model has focused on another aspect, that
of periodic crystal formation.  In this language the A sites are ions, and
the B sites are empty sites.  The question is whether or not the interaction
of the ions with the electrons, coupled with
the fact that the electrons satisfy the
Pauli exclusion principle, is sufficient to cause a periodic arrangement of the
ions to be the ground-state configuration.  This is the
language that we adopt in this contribution.

The Falicov-Kimball model is probably
the simplest many-body physics problem.  The many-body-physics aspect of the
problem enters when one considers an {\it annealed} average over all possible
ion configurations $\{w_i\}$.  At zero temperature, one searches for the
ion configuration that minimizes the energy of the electrons subject to the
relevant constraints that are applied to the system.

The Hamiltonian in Eq.~(\ref{eq: H_fk}) exhibits two different 
symmetries:\cite{lieb_fk} an ion-occupied--empty-site symmetry and an
electron-hole symmetry.  The first symmetry relates the ground-state energy 
(per site) for the configuration of ions 
$\{w_i\}$ to the energy of the conjugate configuration $\{w_i^*\}:=\{1-w_i\}$
\begin{equation}
E_{gs}(U,\rho_e,\{w_i^*\})=E_{gs}(-U,\rho_e,\{w_i\})-U\rho_e\quad ,
\label{eq: ion_sym}
\end{equation}
while the second symmetry employs the unitary transformation $c_i\rightarrow
(-1)^id_i^{\dagger}$ that changes electrons to holes, yielding
\begin{equation}
E_{gs}(U,1-\rho_e,\{w_i\})=E_{gs}(-U,\rho_e,\{w_i\})-U\rho_i\quad .
\label{eq: el_sym}
\end{equation}
These two symmetries allow restriction to the region $\rho_e\le \frac{1}{2}$
and $\rho_i\le \frac{1}{2}$, without a loss in generality.

The Falicov-Kimball model has been actively studied in recent years, ever since
Kennedy and Lieb\cite{lieb_fk} and Brandt and Schmidt\cite{brandt_schmidt}
independently proved that the period-two phase is the ground state for
all $U$ when the electron and the ion densities are both equal to $\frac{1}{2}$.
Most emphasis has concentrated on the one-dimensional model, where numerical
studies\cite{freericks_falicov} indicated that the system phase separated
into the segregated phase (where all the ions cluster on one side of the
lattice) for large enough interaction strength if $\rho_e\ne\rho_i$ or $\rho_e+
\rho_i\ne 1$.  In the other cases, where $\rho_e=\rho_i$ (the neutral case) and
$U\rightarrow \infty$, or $\rho_e+\rho_i=1$ (the mixed-valence case) and $U
\rightarrow -\infty$, it was conjectured that the most homogeneous phase was
the ground state.  These two conjectures have already been proven to be
true\cite{gruber_seg,brandt_seg,lemberger}.  Another conjecture, 
based upon the many-body
version of Rayleigh-Schroedinger perturbation theory, stated that in the 
small $U$ limit the ground-state configuration will be the configuration
that produces the largest gap at the Fermi level, and this state was shown
to have the smallest periodicity that could produce a gap at the Fermi
level (consistent with the Peierls picture).  Recent
analytical\cite{gruber_molecule} and numerical\cite{gruber_farey} work on the
neutral case has 
shown, however, that at low electron density, there is a tendency for molecule
formation, rather than a homogeneous distribution of the ions, and a
phase-separated configuration of ions may yield a lower energy than a pure
periodic phase. 

In the spirit of the nearly free electron model, we establish two rigorous
results which are valid for $U$ sufficiently small:  

First, we show that if the electron density is $\rho_e=\frac{p}{q}$ with $p$
relatively prime to $q$, and $\rho_i=\frac{p_i}{q_i}$, 
with $\frac{p^{\prime}}{q}<\rho_i
<\frac{p^{\prime}+1}{q}$ for some integer $p^{\prime}$, then the ground-state
configuration is a phase-separated mixture of period-$q$ phases, and possibly 
the empty (or full) lattice.

The second result is a statement about the stability of the pure period-$q$
phase for $\rho_e=\frac{p}{q}$ 
(with $p$ relatively prime to $q$) and $\rho_i=\frac{p_i}{q}$.
For $\rho_i\in [\rho_c,1-\rho_c]$ with $\rho_c\approx 0.371$ [solution
of Eq.~(\ref{eq: transcend})], the ground state has period $q$ and is the 
most homogeneous
configuration; it also has the smallest periodicity needed to produce a gap
at the Fermi level.  On the other hand, if the ion density 
$\rho_i=\frac{p_i}{q}$ is 
smaller than $\frac{1}{4}$ or greater than $\frac{3}{4}$, then the ground state
is always a phase-separated mixture of a phase with $\rho_e=\frac{p}{q}$, 
$\rho_i^{\prime}=0$ (or $\rho_i^{\prime}=1$) and a period-$q$ phase with $\rho_e
=\frac{p}{q}$, 
$\rho_i^{\prime\prime}=\frac{p_i^{\prime\prime}}{q}$ a rational that is 
closest to $\rho_c$ in a
well-defined sense.  For $\rho_i=\frac{p_i}{q}<\rho_c$ or  $\rho_i=\frac{p_i}{q}
>1-\rho_c$, the same is true, i.~e. the ground state is a phase-separated
mixture, except for special values of $\rho_i$ [those satisfying 
Eq.~(\ref{eq: p_stability})]
for which the period-$q$ phase is stable.

These results show that the close analogy with the Peierls instability is valid
only for $\rho_c<\rho_i<1-\rho_c$.  We view the analogy as follows:  For $U=0$
(and $\rho_e=\frac{p}{q}$, $\rho_i=\frac{p_i}{q}$ fixed) 
any ion configuration is a ground
state, i.~e. the probability to find an ion at a given site is uniform and 
equals $\rho_i$.  This uniform-density state is the ``undistorted state'',
has no gap in the electronic spectrum, and is metallic.  For $U\ne 0$, 
sufficiently small, a particular ion configuration is selected which has
period $q$.  It corresponds to the Peierls-Fr\" olich ``distorted state'', 
which has a gap at the Fermi level and is insulating.
For $\rho_i<\rho_c$ or $\rho_i>1-\rho_c$, the ground state is (in general)
phase separated and is a mixture of a metallic and an insulating state.  This
situation does not have a counterpart in the standard theory of Peierls and
Fr\" ohlich.

Finally, the above results establish the existence of a phase transition in the
ground state of the Falicov-Kimball model when $U$ is varied.  For densities
such that the ground state is a phase-separated mixture (for $U$ sufficiently
small) there must be a phase transition as $U$ increases.  Indeed, for $U$ 
sufficiently large, the ground state is known to be either the most-homogeneous
phase or the segregated phase (which is a different phase-separated state).

Our presentation is organized as follows:  in Section II the perturbation theory
is developed showing the $U^2\ln U$ behavior of the ground-state energy for
small $U$;  in Section III the perturbation-theory results are analyzed to
show when pure phases are the ground state and when the ground state is
phase-separated; a discussion follows in Section IV.

\section{Perturbation Theory}

It is most convenient to rewrite the Falicov-Kimball Hamiltonian in a 
momentum-space representation before developing a perturbation-series
expansion for the ground-state energy.  Using the standard Fourier
transform
\begin{equation}
a_k:=\frac{1}{\sqrt{N}}\sum_{j=1}^N e^{-ikj}c_j\quad ,
\label{eq: akdef}
\end{equation}
(with the lattice spacing set equal to 1) yields
\begin{equation}
H=\sum_k [\epsilon (k)-UW(0)] a_k^{\dagger}a_k -U\sum_{k\ne k^{\prime}}
W(k-k^{\prime}) a_k^{\dagger}a_{k^{\prime}}\quad ,
\label{eq: H_mom}
\end{equation}
for the Hamiltonian of the Falicov-Kimball model in momentum space.  The
wave vectors $k$ and $k^{\prime}$ are restricted to the first Brillouin
zone $(-\pi<k\le \pi)$
 and $\epsilon (k):=-2t\cos k$ is the unperturbed band-structure.
$W(2\pi n/Q)$ is the structure factor of the period-$Q$ ion configuration
$\{w_i\}$
\begin{equation}
W(2\pi n/Q):= \frac{1}{Q}\sum_{j=1}^{Q}e^{-i(2\pi nj/Q)}w_j\quad ,
\label{eq: wdef}
\end{equation}
defined for $n=0,1,...,Q-1$. (It is notationally simpler here to define the
$k$-vectors with $k=2\pi n/Q$ to sometimes lie outside of the first
Brillouin zone.  Of course, translation by $-2\pi$ will shift these
vectors back into the first Brillouin zone.)
Note that $W(0)=\rho_i$ by definition.

We begin by performing the many-body version of Rayleigh-Schroedinger
perturbation theory with the double-summation term in Eq.~(\ref{eq: H_mom})
acting as the perturbation.  The analysis is 
straightforward\cite{freericks_falicov}, requiring a momentum-space integral
that can be evaluated analytically, yielding
\begin{eqnarray}
E_{gs}(U,\rho_e,\{w_i\})&=&-\frac{2t}{\pi}\sin(\pi \rho_e)-U\rho_e\rho_i\cr
&+&\frac{U^2}{8\pi t}\sum_{n=1}^{Q-1}\frac{|W(2\pi n/Q)|^2}{\sin(\pi n/Q)}\ln
\left | \frac{\sin(\pi n/Q)-\sin (\pi \rho_e)}{\sin(\pi n/Q)+\sin (\pi \rho_e)}
\right | +O(U^3)\quad ,
\label{eq: rs_mb}
\end{eqnarray}
for the ground-state energy of configuration $\{w_i\}$.
 
The perturbative expansion in Eq.~(\ref{eq: rs_mb}) has a singularity when
the electron density is rational $\rho_e=\frac{p}{q}$ and the ion configuration
has a period that is a multiple of $q$ (with the exception of those ion
configurations whose relevant structure factor vanishes).  It was argued 
heuristically in Ref.~\onlinecite{freericks_falicov} that the configuration
with the maximal singularity (i. e., with the maximal value of 
$|W(2\pi\rho_e)|$) will be the ground-state configuration, and this result
agreed with the numerical work.  However, such logic is flawed, because the
expansion in Eq.~(\ref{eq: rs_mb}) is valid for
$U/t\ll |\ln|\sin(\pi n/Q)-\sin (\pi \rho_e)||$, which cannot hold when
an integral number of electronic subbands are filled (i. e., when $\rho_e=
\frac{p}{q}$).  This result
was known by Fr\"ohlich\cite{frohlich}, and it arises from the fact that
there are degeneracies in the unperturbed wavefunction that were neglected
in the above analysis. 

It is easiest to see the origin of the degeneracies and how to properly treat
them by examining the perturbation theory of the single-particle energy
levels.  Wigner-Brillouin perturbation theory is used, because it automatically
removes the singularities.  The ground-state energy is found by simply filling
up the lowest available single-particle energy levels in the system.  These
energy levels can be expanded in a perturbation series which yields
\begin{equation}
E(k,U,\{w_i\})=\epsilon(k)+\frac{U^2}{t}\sum_{n=1}^{Q-1}
\frac{|W(2\pi n/Q)|^2}{E(k,U,\{w_i\})-\epsilon(k+\frac{2\pi n}{Q})}\quad ,
\label{eq: wb_qp}
\end{equation}
to second order in $U$.
The quasiparticle energy $E(k,U,\{w_i\})$ appears on both sides of 
Eq.~(\ref{eq: wb_qp}) because  one must self-consistently solve for the energy
in a Wigner-Brillouin perturbation-theory expansion.  The
equivalent Rayleigh-Schroedinger expansion would replace $E(k)$ by
$\epsilon (k)$ in the right hand side of (\ref{eq: wb_qp}) which produces
a singularity when $k=-\pi n/Q$ because $\epsilon (k)=\epsilon (-k)$.

At this point, textbooks note that the dominant term in the sum over $n$,
in the right-hand side of Eq.~(\ref{eq: wb_qp}), is the term where 
$k+\frac{2\pi n}{Q}$ is closest to $2\pi-k$, i.~e. it is the term with
$n$ closest to $Q(1-k/\pi )$.
If the other terms are neglected, then Eq.~(\ref{eq: wb_qp}) reduces to
a quadratic equation that can be solved exactly.
This procedure is sometimes called nearly degenerate
perturbation theory because it produces the correct secular equation in
the degenerate case. 

However, we choose to proceed in a more rigorous manner in the case where the
value of the interaction is much smaller than the subband width
$U\ll \pi t/Q$.  In this case, the effect of the 
additional terms can be treated in a perturbative fashion, which gives
\begin{eqnarray}
E(k,U,\{w_i\})&=&-t[\cos k+ \cos (k-\frac{2\pi n}{Q})]+\frac{U^2}{t}f_n(k)\cr
&\pm &\sqrt{\{t[\cos k- \cos (k-\frac{2\pi n}{Q})]-\frac{U^2}{t}f_n(k)\}^2+
\frac{U^2}{t}|W(\frac{2\pi n}{Q})|^2} +O(U^3)\quad ,
\label{eq: qp}
\end{eqnarray}
with
\begin{equation}
f_n(k):=-\frac{1}{4}\sum_{{m=1}\atop {m\ne n}}^{Q-1}
\frac{|W(\frac{2\pi m}{Q})|^2}{\cos k -\cos (k-\frac{2\pi m}{Q})}\quad ,
\label{eq: f_def}
\end{equation}
for $\pi (n-\frac{1}{2})/Q < k < \pi (n+\frac{1}{2})/Q$.  The minus sign is
for the subband energy with $k\rightarrow\pi n/Q$ from below, and the
plus sign is for $k\rightarrow\pi n/Q$ from above.  This form for the 
quasiparticle energies is {\it exact for all $U$} 
when $Q=2$, but is perturbative for all higher periods.

The ground-state energy is found by summing up all of the quasiparticle
energies with $|k|<k_F=\pi\rho_e$ ($k_F$ is the Fermi wavevector). Since the 
quasiparticle energies reproduce
the noninteracting result when $U=0$, the zeroth and first-order terms are
correctly produced by this summation.  We want to concentrate on the 
higher-order terms.  The solution for the quasiparticle energies reveals
that a generic period-$Q$ configuration will break into $Q$ subbands.
The band gaps are equal to $2U|W(\frac{2\pi n}{Q})|$ and are symmetrically
displaced
to lowest order; the order $U^2$ correction leads to asymmetries in the
subband-structure.  If the Fermi energy lies within a subband, then it is
easy to show that for $U\ll \pi t/Q$ the shift in the ground-state energy
is of order $U^2/t$, because the square-root in Eq.~(\ref{eq: qp}) can always be
expanded in a convergent
power series in $U$.  However, no such perturbation-series
expansion can be made if the Fermi energy
lies within one of the band gaps.  In this
case the ground-state energy actually has a $U^2\ln U$ 
dependence\cite{frohlich} which
is always larger than any order $U^2$ dependence for small enough $U$.

We illustrate the origin of the $U^2\ln U$ terms in the expansion for
the ground-state energy for rational electron densities 
$\rho_e=\frac{p}{q}$
with $p$ relatively prime to $q$.  We consider any ion configuration with
a period $Q$ that is a multiple of $q$.  This guarantees that there will
be a band gap at the Fermi momentum $k_F=\pi\rho_e$.  The ground-state
energy is
\begin{equation}
E_{gs}(U,\rho_e,\{w_i\})=\sum_{|k|<k_F}E(k,U,\{w_i\})=\frac{1}{\pi}
\int_0^{\pi\rho_e}E(k,U,\{w_i\})dk\quad .
\label{eq: egs}
\end{equation}
Since the band gaps are symmetric to lowest order, the effects of the lower
filled subbands cancel, and the $U^2\ln U$ contribution arises entirely
from filling the uppermost subband.  Therefore, the $U^2\ln U$ contribution
comes from the integral
\begin{equation}
I:=-\frac{1}{\pi}\int_{\pi (\rho_e-\frac{1}{2Q})}^{\pi\rho_e}
\sqrt{\{t[\cos k- \cos (k-2\pi\rho_e)]-\frac{U^2}{t}f_p(k)\}^2+
\frac{U^2}{t}|W(2\pi\rho_e)|^2} dk\quad .
\label{eq: idef}
\end{equation}
Use of the identity $\cos k -\cos (k-2\pi\rho_e)=-2\sin \pi\rho_e \sin (k-\pi
\rho_e)$ and shifting the integration range $k\rightarrow -k+\pi\rho_e$
yields
\begin{equation}
I=-\frac{1}{\pi}\int_0^{\pi/2Q}\sqrt{[2t\sin\pi\rho_e\sin k-\frac{U^2}{t}f_p
(\pi\rho_e-k)]^2+U^2|W(2\pi\rho_e)|^2}dk\quad .
\label{eq: i2}
\end{equation}
The $U^2\ln U$ behavior originates from the region near the origin and
$f_p(\pi\rho_e-k)$ does not depend strongly upon $k$ in this region,
so we can approximate the integral by replacing $\sin k\rightarrow k$ and
$f_p(\pi\rho_e-k)\rightarrow f_p(\pi\rho_e)$. The substitution 
$k\rightarrow (U|W(2\pi\rho_e)|\sinh x +\frac{U^2}{t}f_p)/2t\sin\pi\rho_e$
yields an integrable form for $I$ which contains a constant term and
a $U^2\ln U$ term.  The small-$U$ expansion for the ground-state energy
then becomes
\begin{equation}
E_{gs}(U,\rho_e,\{w_i\})=-\frac{2t}{\pi}\sin\pi\rho_e-U\rho_e\rho_i+
\frac{1}{4\pi t}\frac{|W(2\pi\rho_e)|^2}{\sin \pi\rho_e} U^2\ln U+O(U^2)\quad ,
\label{eq: ulnu}
\end{equation}
which contains no $f_p$ dependence.  The above form is only valid for
$U\ll \pi t/Q$.  This perturbative expansion shows that the ground state
will be found by determining the periodic configuration $\{w_i\}$ that
maximizes the square of the structure factor $|W(2\pi\rho_e)|^2$ at twice
the Fermi momentum.  Furthermore, it eliminates all configurations with
periods $Q$ that are not multiples of $q$, since those states only have
a $U^2$ correction to their ground-state energy because the Fermi level
does not lie within a subband gap.  

\section{Phase-separation analysis}

We are interested in finding the ground state of the Falicov-Kimball model
as a function of the electron and ion densities.
The perturbative expansion in Eq.~(\ref{eq: ulnu}) depends on $\rho_e$ in
the {\it zeroth-order} term, which is a {\it convex} function of the electron
density. Therefore, for $U=0$, phase separation can only occur between two
different ion configurations {\it that have the same electron density
$\rho_e$ as the pure phase}.  

Let us examine the effect of the first-order term.  To order $U$, 
Eq.~(\ref{eq: ulnu})
is a concave function of $(\rho_e,\rho_i)$ and thus the ground state will be
a mixture of two phases with densities $(\rho_e^{\prime},\rho_i^{\prime})$
and $(\rho_e^{\prime\prime},\rho_i^{\prime\prime})$.  We set $\rho_e=\alpha
\rho_e^{\prime}+(1-\alpha)\rho_e^{\prime\prime}$, $\rho_i=\alpha\rho_i^{\prime}+
(1-\alpha)\rho_i^{\prime\prime}$, and $\rho_e^{\prime}=\rho_e+\delta\rho_e$.
Since $-\sin (\pi\rho_e)$ is convex, then for $U=0$, we have
$\delta\rho_e=0$ and the
probability that a given site is occupied by an ion is $\rho_i$.  Hence,
$\delta\rho_e$ tends to zero as $U\rightarrow 0$.  Furthermore, one can check 
that the minimum of 
\begin{equation}
\alpha E(\rho_e^{\prime},\rho_i^{\prime})+(1-\alpha)E(\rho_e^{\prime\prime},
\rho_i^{\prime\prime})\quad ,
\label{eq: usmallseg}
\end{equation}
(at first order) is attained for
\begin{equation}
\delta\rho_e=\frac{U\rho_i}{2t\sin (\pi\rho_e)}\quad ,\quad\quad 
\rho_i^{\prime\prime}=0\quad ,
\label{eq: delta_rhoe}
\end{equation}
and the decrease in the ground-state
energy is of the order $U^2$.  This is negligible in comparison to the 
$U^2\ln U$ term so one can assume $\delta\rho_e=0$ at this order.

It is the coefficient of
the $U^2\ln U$ term that determines which ion configuration yields the
lowest energy.  Since the electron density is fixed in all candidate
ground-state configurations, the criterion for selecting the ground-state
configuration is to maximize the square of the structure factor
$|W(2\pi\rho_e,\{w_i\})|^2$, including the possibility that phase-separated
mixtures may be needed in the maximization.

The construction of the maximum square structure factor is a straightforward
exercise for each phase $\{w_i\}$.
Consider a rational electron density $\rho_e=\frac{p}{q}$
with $p$ relatively prime to $q$ and a rational ion density $\rho_i=\frac{p_i}
{q_i}$ with $p_i$ relatively prime to $q_i$.  Then  the maximum of $|W|^2$
is achieved with the following period-$Q$ ion 
configuration\cite{freericks_falicov} (with $Q={\rm lcm}\{q,q_i\}=:sq$):
Define the $q$ numbers $r_j$ by
\begin{equation}
(pr_j):= j\quad{\rm mod} q\quad ,\quad j=0,1,...,q-1\quad ,
\label{eq: n_def}
\end{equation}
and set $w_i=1$ for
\begin{eqnarray}
i&=&r_j+mq\quad ,\quad\quad j=0,1,...,n-1\quad ,\quad
m=0,1,...,s-1\quad ,\quad n={\rm int} [q\rho_i]\quad ,\cr
&=&r_n+mq\quad ,\quad\quad m={\rm any~} (Q\rho_i-sn) 
{\rm ~numbers~in~the~set~} \{0,1,...,s-1\}~.
\label{eq: wconst}
\end{eqnarray}
Note that the above construction is not necessarily unique when $s\ne 1$,
but every configuration constructed in such a fashion will have the same
square structure factor (the order $U^2$ corrections to the energy should
split any remaining degeneracies).  It is easy to verify that the above
construction does satisfy $W(0)=\rho_i$ and
\begin{eqnarray}
|W(2\pi\rho_e,\rho_i)|^2&=&\frac{1}{Q^2}\sum_{j,k=1}^{Q}w_jw_k\cos [2\pi\rho_e
\frac{(j-k)}{Q}]\quad ,\cr
&=&\frac{n-q\rho_i+(n-q\rho_i)^2}{q^2}\cr
&+&\frac{1}{2q^2}
\frac{1+(q\rho_i-n-1)\cos (2\pi n/q)-(q\rho_i-n)\cos [2\pi (n+1)/q]}{\sin^2\pi /q}
\quad .
\label{eq: wsquare}
\end{eqnarray}
In the special case where $s=1$, so that
$n=q\rho_i$, the above form simplifies to
\begin{equation}
|W(2\pi\rho_e,\rho_i)|^2=\frac{1}{2q^2}
\frac{(1-\cos 2\pi\rho_i )}{\sin^2\pi /q}=\frac{1}{q^2}
\frac{\sin^2\pi\rho_i }{\sin^2\pi /q} \quad \quad (s=1)
\quad .
\label{eq: wsquare_1}
\end{equation}
Note that $|W(2\pi\rho_e,\rho_i)|^2$ depends on $\rho_e$ only through the
denominator $q$.  This fact greatly simplifies the analysis below.

The ion configuration that maximizes the square of the structure factor
is identical to Lemberger's most-homogeneous configuration\cite{lemberger}
in the neutral case $\rho_i=\rho_e$.  In the nonneutral cases, the maximal
ion configuration satisfies uniform-distribution 
properties\cite{freericks_falicov} in which the configuration is composed
of clusters of ions, with only islands of size $l$ and $l-1$ appearing.
Furthermore, these islands are ``most-homogeneously''
distributed (the most-homogeneous configuration is the special case with
islands of size 1).

{\it Lemma (local convexity of the squared structure factor).  Assume
the electron density is rational $\rho_e=\frac{p}{q}$ with $p$ relatively
prime to $q$, and assume that the ion density is also rational
$\rho_i=\frac{p_i}{q_i}$ with $p_i$ relatively prime to $q_i$, and satisfies
$\frac{p^{\prime}}{q} < \rho_i < \frac{p^{\prime}+1}{q}$ for some integer 
$p^{\prime}$.
Then a mixture of ionic phases with ion densities 
$\frac{p^{\prime}}{q}$ and $\frac{p^{\prime}+1}{q}$ will have a larger square 
structure factor than the pure phase with ion density $\rho_i$.}

{\it Proof:} We need to show that the maximal square structure factor in
Eq.~(\ref{eq: wsquare}) is locally convex.  To do this we must examine
the condition for convexity, by computing
\begin{equation}
C=(p^{\prime}+1-q\rho_i)|W(2\pi\rho_e,\frac{p^{\prime}}{q})|^2
+(q\rho_i-p^{\prime})|W(2\pi\rho_e,\frac{p^{\prime}+1}{q})|^2
-|W(2\pi\rho_e,\rho_i)|^2\quad .
\label{eq: cdef}
\end{equation}
If $C>0$, then the square structure factor is locally convex, and the lemma
will have been proven.  Substituting Eqs.~(\ref{eq: wsquare}) and
(\ref{eq: wsquare_1}) into Eq.~(\ref{eq: cdef}) yields
$C=[q\rho_i-p^{\prime}+(q\rho_i-p^{\prime})^2]/q^2$ which is greater than
zero for $0<q\rho_i-p^{\prime}<1$, which is a condition that
holds by hypothesis. Q.E.D.

{\it Comment.} The above lemma shows that the search for a maximal square
structure factor can be limited to those ion configurations {\it that
possess the minimal periodicity $q$ needed to produce a gap at the Fermi
level}.  It has not determined the global maximum.  That search will be
completed below.  The lemma does allow us to immediately prove our first
result about phase separation in the Falicov-Kimball model.

{\it Theorem 1 (minimal-period phase separation).  If the electron
and ion densities satisfy the hypothesis of the lemma, then for $U$
sufficiently small (i.~e., $Uq\ll 1)$, the ground-state configuration
is a mixture of two period-$q$ phases, with densities $\rho_i^{\prime}=\frac{
p^{\prime}}{q}$ and $\rho_i^{\prime\prime}=\frac{p^{\prime\prime}}{q}$ 
($p^{\prime}$ or $p^{\prime\prime}$
can be equal to $0$ or $q$).}

{\it Proof:} The perturbative analysis of Section II established that in
the limit $U\rightarrow 0$ the
ground-state configuration is determined by maximizing the 
square of the structure factor evaluated at twice the Fermi wavevector.
The above lemma shows that such a search can be limited to a search over
ion configurations with $\rho_i^{\prime}=\frac{p^{\prime}}{q}, \ 
p^{\prime}=0,...,q$.  
This means that if the ion density does not equal $\frac{p^{\prime}}{q}$,
then it must phase separate into a mixture of states whose
electron densities are $\rho_e=\frac{p}{q}$ and whose ion densities are
$\frac{p^{\prime}}{q}$ and $\frac{p^{\prime\prime}}{q}$. Q.E.D.

It follows from the Lemma that we only need to search for the ground state 
among the ion configurations with period $q$ (given $\rho_e=\frac{p}{q}$ 
with $p$
relatively prime to $q$), therefore $s=1$ and the square structure factor is
given by  Eq.~(\ref{eq: wsquare_1}).

The function $(\cos 2\pi\rho_i-1)$ is concave for $\rho_i\in 
[0,\frac{1}{4}]\cup [\frac{3}{4},1]$ and convex for $\rho_i\in 
[\frac{1}{4},\frac{3}{4}]$.  Hence, if $\rho_i$ lies in the interval
$[0,\frac{1}{4}]\cup [\frac{3}{4},1]$, the pure-phase cannot be stable
against phase separation.

The convex envelope of the function $(\cos 2\pi\rho_i-1)$ is given by
\begin{eqnarray}
(\cos 2\pi\rho_c-1)\frac{\rho_i}{\rho_c}~&{\rm for}&~0\le\rho_i\le\rho_c\cr
(\cos 2\pi\rho_i-1)~&{\rm for}&~\rho_c\le\rho_i\le 1-\rho_c\cr
(\cos 2\pi\rho_c-1)\frac{1-\rho_i}{\rho_c}~&{\rm for}&~1-\rho_c\le\rho_i\le 1
\label{eq: convex_envelope}
\end{eqnarray}
where $\rho_c\approx 0.3710$ is the solution to the equation
\begin{equation}
2\pi\rho_c=\tan \pi\rho_c\quad .
\label{eq: transcend}
\end{equation}
Thus if $\rho_i$ is a rational in the interval $[\rho_c,1-\rho_c]$, the
pure phase with $\rho_e=\frac{p}{q}$, $\rho_i=\frac{p_i}{q}$ is stable.

Let us now analyze what happens for densities $\rho_i$ in the interval
$[0,\rho_c]$.  The case $[1-\rho_c,1]$ is similar.  The Lemma states that we 
must consider only the ion densities in the discrete set $\{\rho_i=\frac
{p^{\prime}}{q}\}$.  Given $\rho_e=\frac{p}{q}$, let $\frac{\tilde p_i}{q}$
be the largest rational in the set $\{\frac{p^{\prime}}{q}\}$ which is smaller
than $\rho_c$.  From the construction of the convex envelope, for any
$\rho_i<\frac{\tilde p_i}{q}$ we know that the ground-state configuration is
a mixture of the empty configuration $\rho_i^{\prime}=0$ and a period-$q$
configuration with density $\rho_i^{\prime\prime}=\frac{\tilde p_i}{q}$ or
$\frac{\tilde p_i+1}{q}$.

To decide between the two possible values of $\rho_i^{\prime\prime}$, we have
to determine whether $\frac{\tilde p_i}{q}$ corresponds to a pure phase,
or a mixture of the empty state and a period-$q$ configuration with
density $\frac{\tilde p_i+1}{q}$. Using Eqs.~(\ref{eq: ulnu}) and 
(\ref{eq: wsquare_1}), it follows that the pure phase $\frac{\tilde p_i}{q}$
is stable if
\begin{equation}
\sin\pi\frac{\tilde p_i}{q}>\left (\frac{\tilde p_i}{\tilde p_i+1}
\right )^{\frac{1}{2}}\sin\pi\frac{\tilde p_i+1}{q}\quad ,
\label{eq: p_stability}
\end{equation}
and unstable if Eq.~(\ref{eq: p_stability}) is not satisfied.

To summarize, given $\frac{p}{q}$ and $\frac{\tilde p_i}{q}$ the largest
rational with denominator $q$ that is smaller than $\rho_c$, if 
Eq.~(\ref{eq: p_stability}) is satisfied, 
then for $\rho_i=\frac{\tilde p_i}{q}$ (resp. 
$1-\frac{\tilde p_i}{q}$) the ground-state configuration is periodic, given by 
Eq.~(\ref{eq: n_def}), and for all $\rho_i<\frac{\tilde p_i}{q}$ the ground
state is a mixture with $\rho_i^{\prime}=0$ and $\rho_i^{\prime\prime}=
\frac{\tilde p_i}{q}$ (resp. for all $\rho_i>1-\frac{\tilde p_i+1}{q}$,
$\rho_i^{\prime}=1$ and $\rho_i^{\prime\prime}=1-\frac{\tilde p_i}{q}$).
On the other hand, if Eq.~(\ref{eq: p_stability}) is not satisfied, then
for all $\rho_i<\frac{\tilde p_i+1}{q}$ the ground state is a mixture
with $\rho_i^{\prime}=0$ and $\rho_i^{\prime\prime}=\frac{\tilde p_i+1}{q}$,
and similarly for $\rho_i>1-\frac{\tilde p_i}{q}$.

In Table 1, we give the values of $\tilde p_i$ for $q=3$ to 34, and indicate
whether the pure phase with $\rho_i=\frac{\tilde p_i}{q}$ is stable (s)
or unstable (u).  For example, the state with $\rho_e=\frac{4}{15}$ is unstable
for any $\rho_i<\frac{6}{15}$ or $\rho_i>\frac{9}{15}$ and stable for
$\rho_i=\frac{6}{15}$, $\frac{7}{15}$, $\frac{8}{15}$, $\frac{9}{15}$.  The
state with $\rho_e=\frac{2}{9}$ is unstable for $\rho_i<\frac{3}{9}$ or
$\rho_i>\frac{6}{9}$ and stable for $\rho_i=\frac{3}{9}$, $\frac{4}{9}$,
$\frac{5}{9}$, $\frac{6}{9}$.  In these two examples the neutral state
$\rho_e=\rho_i$ is unstable.  On the other hand, for $\rho_e=\frac{4}{11}<
\rho_c$, the neutral state is stable.

In general, the neutral state $\rho_e=\rho_i$ is unstable for $\rho_i<\rho_c$,
with an infinite number of exceptions [given by Eq.~(\ref{eq: p_stability})]
for which the first few electron densities are $\rho_e=\frac{1}{3}$,
$\frac{1}{4}$, $\frac{4}{11}$, $\frac{5}{14}$, $\frac{6}{17}$, $\frac{7}{19}$,
$\frac{7}{20}$, $\frac{9}{25}$, $\frac{10}{27}$, $\frac{11}{30}$.  The state
with diatomic molecules $\rho_i=2\rho_e$ is unstable for $\rho_i<\rho_c$
with an infinite number of exceptions $\rho_e=\frac{1}{6}$, $\frac{2}{11}$,
$\frac{3}{17}$, $\frac{5}{27}$, $\frac{5}{28}$, $\frac{7}{38}$, \ldots
Similarly, in the triatomic case $\rho_i=3\rho_e$, the exceptional
electronic densities for which the pure state is stable are $\rho_e=
\frac{1}{9}$, $\frac{2}{17}$, $\frac{3}{25}$, $\frac{4}{33}$, \ldots
In any case, it appears that for any $\epsilon>0$ and for any state with
$n$-molecules $\rho_i=n\rho_e$ there is a finite number of exceptions
in $[\frac{1}{4},\rho_c-\epsilon]$ as shown in Figure 1.

These results lead us to the theorem:

{\it Theorem 2.  If the electron density is rational, $\rho_e=\frac{p}{q}$ with
$p$ relatively prime to $q$, and the ion density is $\rho_i=\frac{p_i}{q}$, then
\begin{description}
\item [a)] 
for $\frac{p_i}{q}\in [\rho_c,1-\rho_c]$, or $\frac{p_i}{q}=\frac{\tilde
p_i}{q}$ with $\tilde p_i$ solving Eq.~(\ref{eq: p_stability}), the ground-state
configuration is periodic with period $q$.
\item [b)]
for $\frac{p_i}{q}<\rho_c$ (or $\frac{p_i}{q}>1-\rho_c$) and $\frac{p_i}{q}
\ne\frac{\tilde p_i}{q}$ with $\tilde p_i$ solving Eq.~(\ref{eq: p_stability}),
the ground-state configuration is a mixture of the empty lattice 
$\rho_i^{\prime}=0$ and the period-$q$ configuration with $\rho_i^{\prime\prime}
=\frac{\tilde p_i+1}{q}$, (resp. $\rho_i^{\prime}=1$ and $\rho_i^{\prime\prime}
=1-\frac{\tilde p_i+1}{q}$).
\item [c)]
for all $\frac{p_i}{q}<\frac{1}{4}$ or $\frac{p_i}{q}>\frac{3}{4}$, the ground
state configuration is a mixture like in b).
\end{description}
}

{\it Comments}: (i)
The exceptional ion densities can all be found by studying
Eq.~(\ref{eq: p_stability}).  We have not
been able to determine an explicit formula for these exceptional ion
densities. (ii) The phase-separated state is not an insulating state, but
rather is the mixture of a metallic state (the empty lattice) and an insulating
state (the period-$q$ phase with $p$ filled subbands).
(iii) In the neutral case, $\rho_e=\rho_i=\rho$, for any $\rho\in
[\rho_c,1-\rho_c]$ and for the ``exceptional'' values in the intervals $[0,
\rho_c]$ or $[1-\rho_c,1]$, the ground state is most homogeneous, since
the state with the maximal structure factor satisfies the uniform-distribution
property.  It is also the configuration obtained by Lemberger's 
construction\cite{lemberger}.  For these pure states it is expected that the
ground state does not have any phase transition when $U$ increases from $+0$ 
to $+\infty$, since for any rational density $\rho$ the ground state is known to
be the most homogeneous state for $U$ sufficiently large.  This expectation is 
also confirmed for intermediate values of $U$ $(U\ge 0.1)$ by exact numerical
calculations\cite{gruber_farey}.  Using the same argument for the
``regular'' values of $\rho$ in $[0,\rho_c]$ or $[1-\rho_c,\rho_c]$, there 
will be a phase transition as $U$ varies. (iv) These results only hold for
$U$ sufficiently small with respect to $\frac{1}{q}$, where 
$\rho_e=\frac{p}{q}$, for the $U^2\ln U$ term to dominate the perturbation 
expansion.  For $\rho_e=\frac{p}{q}$ and $\rho_i=\frac{p_i}{q}\ne\rho_e$ the 
phase separation that may occur for small $U$ rapidly disappears as $U$ is
increased from 0 to $\infty$ to yield either a pure state or the
segregated phase\cite{gruber_farey}.  For $U$ sufficiently large, it is expected
that the state is either neutral or the segregated phase.

\section{Conclusion}

The band theory of solids is perhaps the defining theory for condensed-matter
physics.  It has been applied to virtually every interesting material that
has been studied.  Nevertheless, the conventional wisdom of Peierls and
Fr\"ohlich for optimizing the band-structure  for the  ground-state of 
one-dimensional crystals
is not always correct.  They argue, that the ion configuration that
produces the largest gap at the Fermi level will yield the ground state.
We find that this argument is true for a nearly free electron model
only if the ion density is close enough to 
half-filling.  For ion densities away from half-filling, the system will phase
separate into a mixture of states that have the same electron density,
but have different ion densities ($\rho_i=0$ and $\rho_i$ close to 0.371
or $\rho_i=1$ and $\rho_i$ close to 0.629).
It is possible that this phase separation can be observed in 
quasi-one-dimensional metals and insulators.  We are not aware of any
experiments that have seen this phase separation.

Our rigorous results hold only for $U$ sufficiently small because they
are based on perturbation-theory arguments that maximize the leading
corrections of the ground-state energy as a function of $U$.  Since these
corrections of order $U^2\ln U$ will compete with order $U^2$ corrections
for finite values of $U$, the phase separation discovered here may rapidly
disappear as $U$ increases. Numerical evidence indicates that this is true for
the densities between $\frac{1}{4}$ and $\frac{3}{4}$, but larger values of $U$
are necessary for the densities near 0 or 1.

Furthermore, since the ground state is known to
be either a different phase-separated state (nonneutral cases) or the 
most-homogeneous state (neutral case) for large $U$, the spinless 
Falicov-Kimball model must have a phase transition as a function of $U$.
In the neutral case, when the ground state is not a phase-separated state,
but is the Peierls-type state that maximizes the band gap at the Fermi level,
it is possible that the ground state has no phase transitions for $0<U<+\infty$,
since the small-$U$ ground state is identical to the large-$U$ ground
state.  We are unable to prove this conjecture here.

Our analysis was restricted to the spinless Falicov-Kimball model, but the
general ideas may also hold for more complicated models such as tertiary
alloy problems (where $w_i$ would assume three different values) or the static
Holstein model (where $w_i$ is continuous), but the determination of the
maximal structure factor becomes much more complicated, since one must maximize
with respect to both the phase and the amplitude, as opposed to maximizing 
only with respect to the phase, as we did here.

\acknowledgments  
We would like to acknowledge useful conversations with E. H. Lieb and D.
Ueltschi.  J.~K.~F. would like to acknowledge the hospitality of the
Institut de Physique Th\'eorique at the EPFL, where this work was
started in June of 1995.  J.~K.~F. would also like to acknowledge the
Donors of The Petroleum Research Fund, administered by the
American Chemical Society, for partial support of this research 
(ACS-PRF\# 29623-GB6).

\mediumtext
\begin{table}
\caption{
Largest integer $\tilde p_i$ such that $\frac{\tilde p_i}{q}<\rho_c
\approx 0.371$. The letters s and u denote whether Eq.~(\ref{eq: p_stability})
is satisfied (s), implying the pure phase $\rho_i=\frac{\tilde p_i}{q}$ is
stable, or is not satisfied (u), implying the pure phase
$\rho_i=\frac{\tilde p_i}{q}$ is unstable.}
\label{table1}
\begin{tabular} {ccccccccccccccccccccccccccccccccc}
$q$&3&4&5&6&7&8&9&10&11&12&13&14&15&16&17&18&19&20&21&22&23&24&25&26&27&28&29&30&31&32&33&34\\
\tableline
$\tilde p_i$&1&1&1&2&2&2&3&3&4&4&4&5&5&5&6&6&7&7&7&8&8&8&9&9&10&10&10&11&11&11&12&12\\
 &s&s&u&s&u&u&s&u&s&s&u&s&u&u&s&u&s&s&u&s&u&u&s&u&s&s&u&s&u&u&s&u\\
\end{tabular}
\end{table}

\begin{figure}[t]
\caption{Stable periodic configuration for $n$-molecules, i.~e. states with
$\rho_i=n\rho_e$.  The values $n=1$ (solid dot), $n=2$ (open square), $n=3$
(solid triangle), $n=4$ (open dot), $n=5$ (solid square), $n=6$ (open triangle),
and $n=7$ (x) are all plotted.  The phases are stable above $\rho_c$ as
indicated by the solid lines.  The dashed lines are guides to the eye.}
\end{figure}

\begin{figure}
\epsfxsize=6.0in
\epsffile{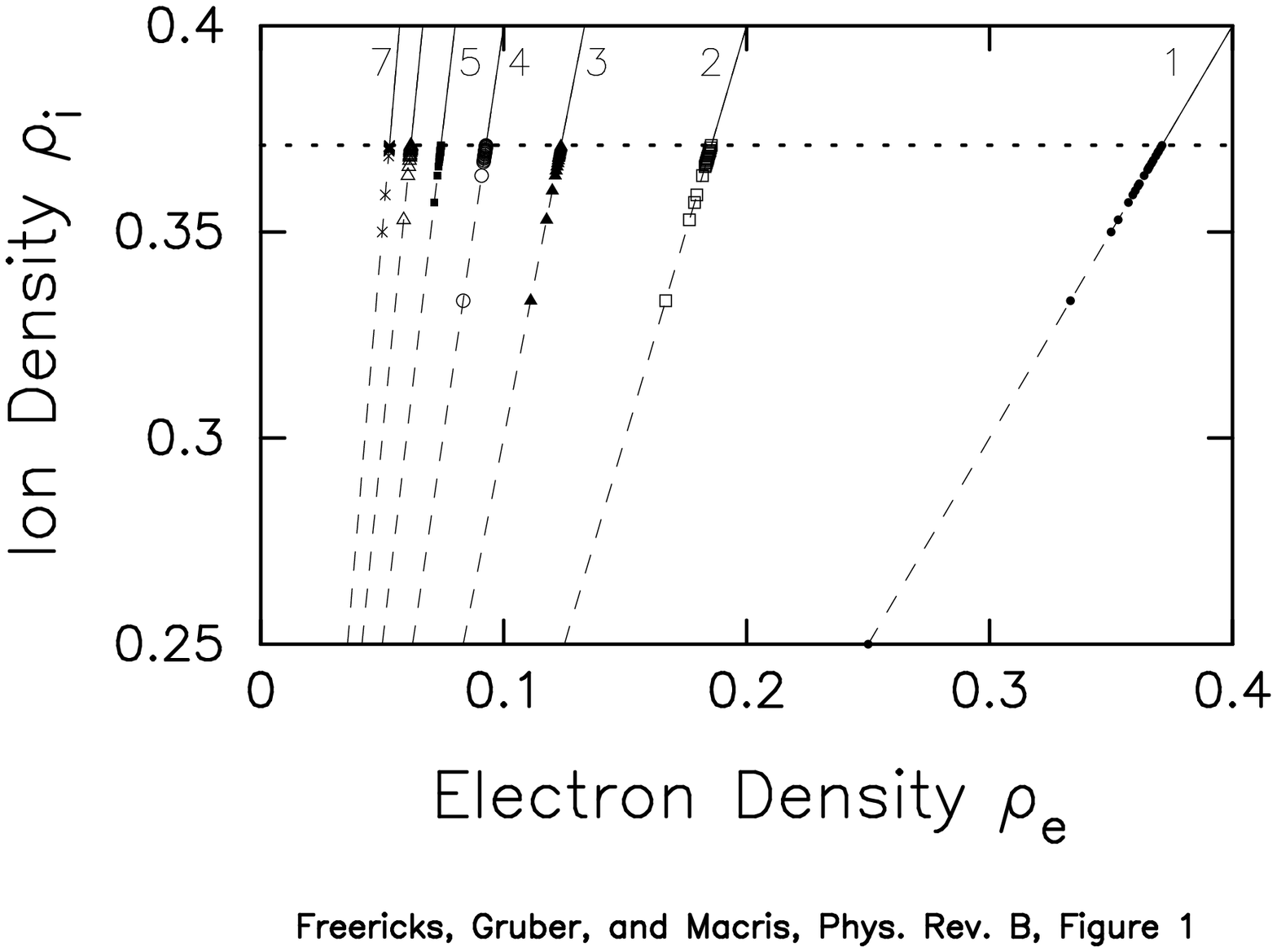}
\end{figure}

\end{document}